\documentclass[prl,superscriptaddress,twocolumn]{revtex4}

\usepackage{graphicx,color}
\usepackage{amsmath}
\usepackage{dsfont}
\usepackage{amssymb}
\usepackage{hyperref}

\newcommand\bwt         {\begin{widetext}}
\newcommand\ewt         {\end{widetext}}

\begin{document}

\title{Geometrically protected triple-point crossings in an optical lattice}

\author{I. C. Fulga}
\affiliation{Institute for Theoretical Solid State Physics, IFW Dresden, 01171 Dresden, Germany.}

\author{L. Fallani}
\affiliation{Department of Physics and Astronomy, University of Florence \& LENS European Laboratory for Nonlinear Spectroscopy, 50019 Sesto Fiorentino, Italy}

\author{M. Burrello}
\affiliation{Niels Bohr International Academy and Center for Quantum Devices, University of Copenhagen, Juliane Maries Vej 30,
2100 Copenhagen, Denmark.}

\begin{abstract}
We show how to realize topologically protected crossings of three energy bands, integer-spin analogs of Weyl fermions, in three-dimensional optical lattices. Our proposal only involves ultracold atom techniques that have already been experimentally demonstrated and leads to isolated triple-point crossings (TPCs) which are required to exist by a novel combination of lattice symmetries. The symmetries also allow for a new type of topological object, the type-II, or tilted, TPC. Our Rapid Communication shows that spin-1 Weyl points, which have not yet been observed in the bandstructure of crystals, are within reach of ultracold atom experiments.
\end{abstract}

\maketitle

\emph{Introduction}.
In the past decades the theory of condensed matter witnessed a topological revolution, sparked by the study of quantum Hall systems \cite{Klitzing1980} and consolidated by the discovery of topological insulators. Noninteracting gapped systems have been classified based on the topological properties of their energy bands and on the symmetries of the underlying Hamiltonians \cite{Hasan2010,Qi2011,Bernevig2013book}, first considering non-spatial symmetries only \cite{Schnyder2008} and then including also crystal symmetries \cite{Chiu2016}. This classification of symmetry-protected topological phases of matter has been recently extended to gapless systems \cite{Chiu2016}, which provide the possibility of simulating many of the particles appearing in high-energy quantum field theories, such as Dirac, Majorana, and Weyl fermions.
The latter appear in a condensed matter setting as band-touching points with a linear dispersion \cite{Lv2015,Lv2015a,Xu2015,Xu2015a} characterized by a 
low-energy Hamiltonian of the form $H_{\rm Weyl}=\vec{k}\cdot\vec{\sigma}$, where $\vec{k}$ is the crystal momentum and $\vec{\sigma}$'s are three Pauli matrices parametrizing the degree of freedom associated with the two touching bands. These degeneracy points are topological defects of nonzero chirality, which gives rise to many interesting phenomena, such as gapless surface Fermi arcs \cite{Wan2011} and the chiral anomaly \cite{Nielsen1983}.

The realization of similar quasiparticles in condensed-matter systems, however, is not bound by the Poincar\'e symmetry of quantum field theories but by the symmetries of the space group describing the lattice. This enables the possibility for designing even more exotic objects, which transcend the usual constraints of the elementary particles \cite{Heikkilae2015, Hyart2016, Weng2016, Zhu2016, Winkler2016, Fulga2017, Lepori2017, Yang2017}. Some of them, for example, carry a multiple chirality and are characterized by nonlinear dispersion relations \cite{Fang2012, Huang2016, Lepori2016}. In most cases, however, the stability of these novel band crossings requires the presence of one or more protecting lattice symmetries. Without them the topological defect would either be destroyed or decomposed into multiple simpler constituents.

Recently it has been shown that a new type of chiral fermion can be realized in lattice models with specific symmetries. It is formed by three energy bands touching at a single point in the Brillouin zone (BZ), a triple-point crossing (TPC), which can be interpreted as a spin-1 fermion \cite{Bradlyn2016}. Two of the energy bands disperse linearly in all momentum directions, whereas the third is locally flat, leading to an effective low-energy Hamiltonian of the form $H_{\rm TPC}=\vec{k}\cdot\vec{S}$. Now, in place of the spin-$\frac{1}{2}$ Pauli matrices describing Weyl cones, the three $3\times3$ matrices $(S_j)_{kl} =-i\varepsilon_{jkl}$ represent an effective pseudospin-1 degree of freedom associated with the three bands. The TPC forms a double monopole of the Berry curvature that results in the appearance of chiral surface states or Fermi arcs. The latter emerge in pairs from the surface projections of the bulk band touching points.

So far, several works have investigated the appearance and properties of TPCs. They have been proposed to exist in the BZ of Pd$_3$Bi$_2$S$_2$ slightly above the Fermi level as well as in Ag$_3$Se$_2$Au \cite{Bradlyn2016, Luo2016}. Spin-1 Weyl points have also been predicted in the phonon spectra of transition-metal monosilicides, such as FeSi \cite{Zhang2018}. Very recently, microwave-assisted level transitions in a superconducting transmon were experimentally employed to simulate TPCs \cite{Tan2017}.

In a cold atom setting, it was suggested that TPCs may be obtained from the Hamiltonian describing a Weyl semimetal by directly replacing the Pauli $\vec{\sigma}$ matrices with spin-1 $\vec{S}$ matrices \cite{Zhu2017,Hu2017}. This approach is both elegant and straightforward but does not clarify which, if any, lattice symmetries are responsible for the stability of TPCs. For instance, it could lead to TPCs which are fine-tuned in the sense that symmetry-allowed perturbations may split the spin-1 fermion into multiple spin-$\frac{1}{2}$ Weyl cones.
Here, we take a different perspective and discuss TPCs which are constrained to appear due to the real-space geometry of the lattice. Hence we refer to them as \textit{geometrically protected} TPCs. 

Our proposal is based on the experimental successes in the quantum simulation of gauge potentials in cold atom systems \cite{Goldman2014}. Ultracold gases trapped in optical lattices offer two ingredients which can be controlled with high accuracy and constitute the basis of our construction: (i) the possibility for introducing nontrivial lattice potentials based on the control over the laser polarizations and interference and
(ii) the introduction of magnetic fluxes generated by Raman lasers, which have proved a viable tool for the simulation of the two-dimensional (2D) Hofstadter model \cite{Aidelsburger2013,Miyake2013,Aidelsburger2014}.

We combine the two above-mentioned elements to engineer a spin-1 Weyl phase of ultracold atoms in a three-dimensional (3D) lattice that is a higher-dimensional generalization of the well-known 2D Lieb lattice.
Symmetry-enforced TPCs appear in the BZ of the system, inheriting their topological protection from the geometrically induced chiral (sublattice) symmetry of the original Lieb lattice. The TPCs form robust double monopoles of the Berry curvature and lead to gapless topological Fermi arcs. We present an optical lattice scheme able to induce this phase of matter in ultracold atomic gases by exploiting only techniques already available in experiments.
In addition, we explain how by modifying this setup one can obtain a new kind of topological object, a tilted spin-1 Weyl, which we dub a \emph{type-II TPC} by analogy with the existing Weyl cone terminology. The type-II crossing does not appear in current condensed-matter proposals because of the multitude of lattice symmetries constraining the shape of the energy bands. Finally, we comment on ways of experimentally accessing the TPCs in our lattice model as well as on directions for future research.

\begin{figure}[tb]
\includegraphics[width=0.95\columnwidth]{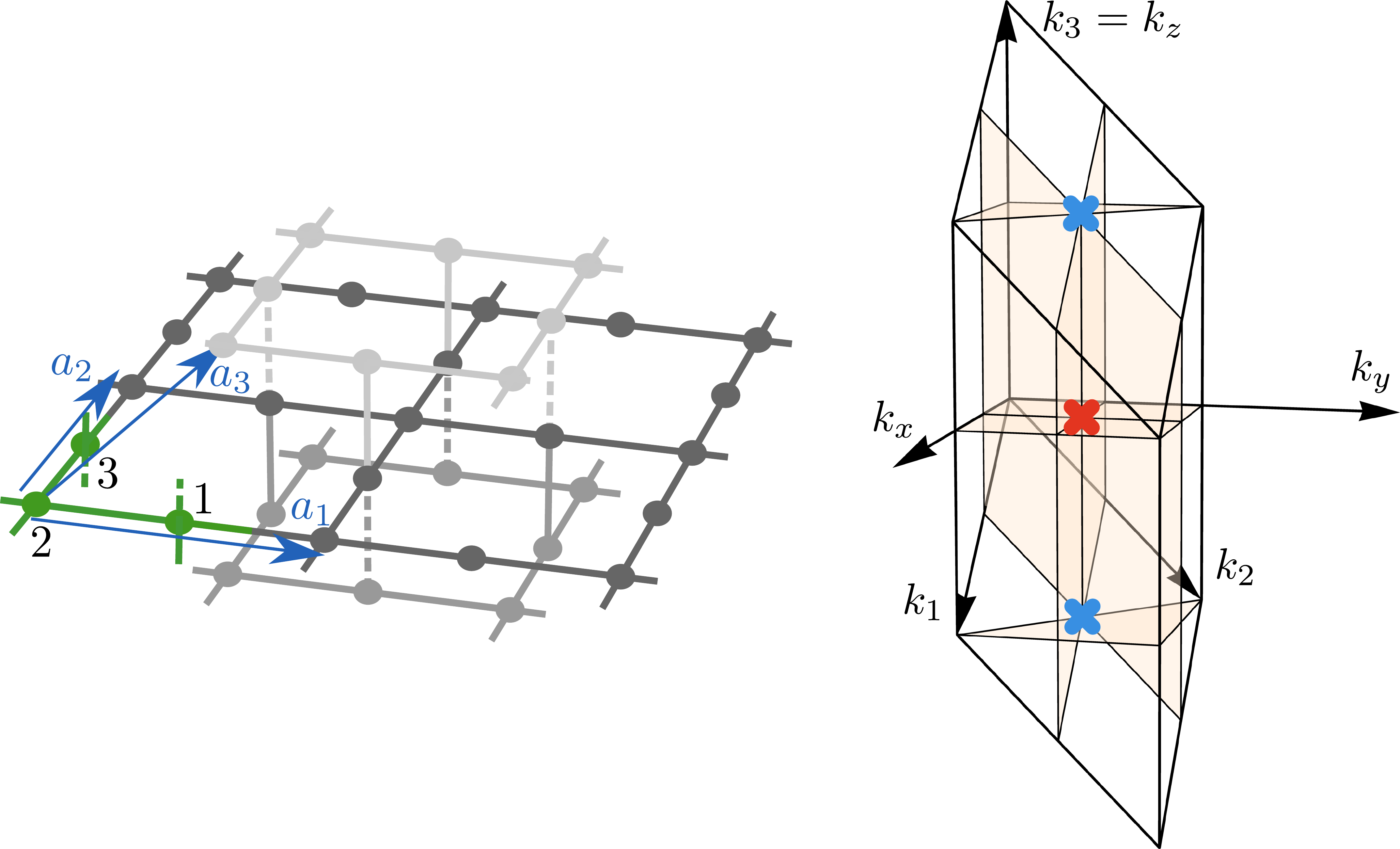}
\caption{Left: A tight-binding model consisting of 2D Lieb lattices stacked in the third dimension. The unit cell (green) consists of three sites labeled 1-3 with Bravais vectors $\vec{a}_1$, $\vec{a}_2$, and $\vec{a}_3$ (blue). Adjacent 2D layers are shifted with respect to each other diagonally. The dashed lines indicate negative hopping amplitudes. Right: Brillouin zone of the model, indicating the momentum vectors as well as the four chiral planes (light orange) located at $k_x=\pi/2$, $k_y=\pi/2$, $k_z=0$, and $k_z=\pm\pi$. Two TPCs (the red and blue crosses) are formed at the triple intersections between the chiral planes, $k_x=k_y=\pi/2$ and $k_z=0,\pi$.\label{fig:system}}
\end{figure}

\emph{Three-dimensional Lieb lattice}.
We consider noninteracting spinless particles hopping on a 3D lattice with a three-site unit cell as shown in Fig.~\ref{fig:system}. The system can be seen as a cubic lattice where the (even, even, even) and (odd, odd, odd) sublattices have been suppressed. The Bravais vectors are $\vec{a}_1=(2, 0, 0)$, $\vec{a}_2=(0, 2, 0)$, and $\vec{a}_3 = (1, 1, 1)$ in units of the inter-site spacing $a$, leading to reciprocal vectors $\vec{k}_1=(\frac{1}{2},0,-\frac{1}{2})$, $\vec{k}_2=(0, \frac{1}{2},-\frac{1}{2})$, and $\vec{k}_3=(0,0,1)$. The main property of this lattice is that each section parallel to the orthogonal planes $xy$, $yz$ and $xz$ is a 2D Lieb lattice, which intuitively justifies the existence of a 3D flat band in its single-particle spectrum. We couple the dynamics of the fermions in the lattice with an artificial vector potential $\vec{A} = \pi/a^2 (0,0,x-z+1/2)$ which corresponds to a synthetic magnetic field along the $\hat{y}$ direction $B_y = -\pi/a^2$. Setting $a=1$, the resulting Hamiltonian is as follows:

\begin{widetext}
\begin{equation} \label{eq:Hlol}
H = - J\begin{pmatrix}
0 & 1 + e^{-2ik_x} & e^{i\left(k_z-k_x+k_y\right)}-e^{-i\left(k_z+k_x-k_y\right)}\\
1 + e^{2ik_x} & 0 & 1 + e^{2ik_y} \\
e^{-i\left(k_z-k_x+k_y\right)}-e^{i\left(k_z+k_x-k_y\right)} & 1 + e^{-2ik_y} & 0 
\end{pmatrix}\,,
\end{equation}
\end{widetext}
where $J$ is the hopping strength and $k_{x,y,z}$ are momenta along the three principal directions. In this gauge, the hoppings in the $\hat{z}$ direction alternate in sign due to the synthetic magnetic field as shown in Fig.~\ref{fig:system}, whereas they remain constant in the $\hat{x}$ and $\hat{y}$ directions.

\begin{figure}[tb]
\includegraphics[width=0.95\columnwidth]{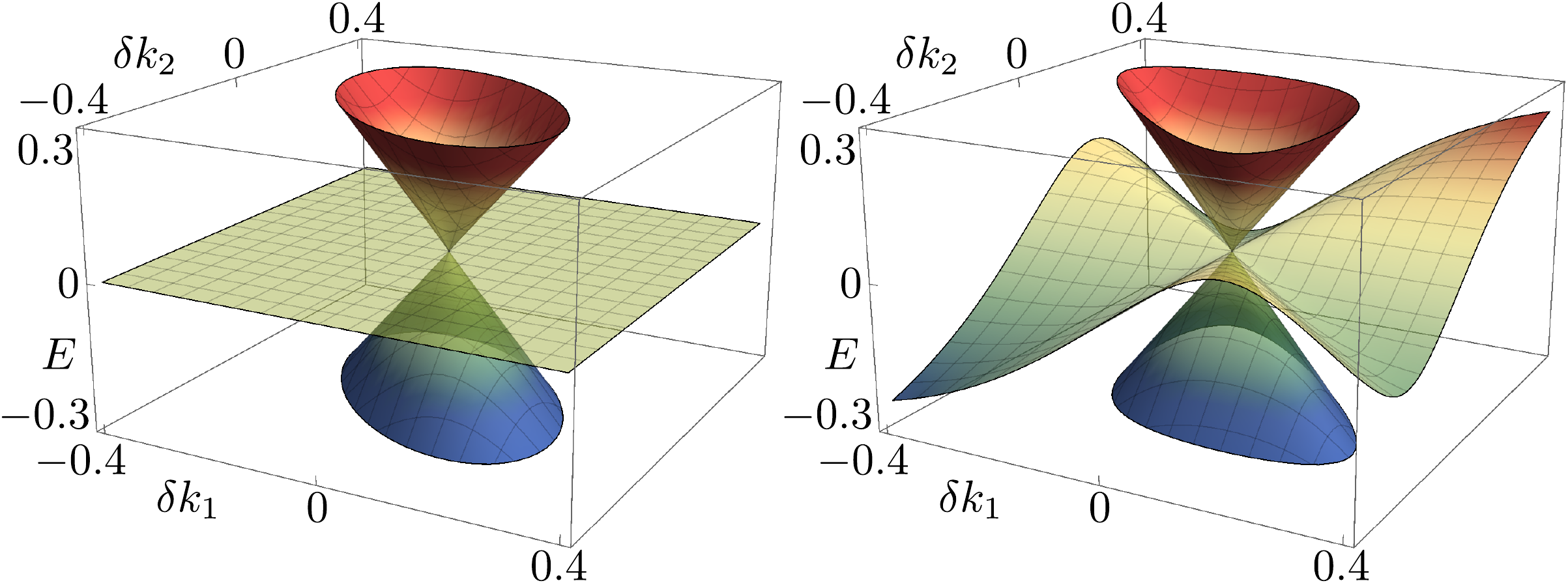}
\caption{Bandstructure of the 3D Lieb lattice close to the TPC position, $k_1=\pi+\delta k_1$, $k_2=\pi+\delta k_2$, and $k_3=0$. Left: In the type-I TPC of Eq.~\eqref{eq:Hlol}, two bands disperse linearly whereas the third remains flat. Right: for the type-II TPC obtained from the Hamiltonian \eqref{eq:Htype2} with $\phi=1$, the central band is tilted, having a nonzero velocity along some momentum directions.\label{fig:tpcband}}
\end{figure}

Diagonalizing Eq.~\eqref{eq:Hlol} yields two TPCs positioned at zero energy and momenta $k_x=k_y=\pi/2$, $k_z=0,\pi$, which have opposite monopole charges. At $k_z=0,\pi$, the effective low-energy Hamiltonian close to the TPC is $H_{0,\pi}=2(-k_x S_3+k_y S_1\pm k_zS_2)$, such that two of the energy bands disperse linearly and the third remains flat (see Fig.~\ref{fig:tpcband}, left panel).

Unlike Weyl cones, which are single monopoles of the Berry curvature and are therefore robust to any sufficiently weak perturbation preserving translational invariance, the TPCs require additional lattice symmetries to remain stable. Here the protecting symmetries are inherited from those of the 2D Lieb lattice. The latter has a chiral (sublattice), $E\to-E$ symmetry, and an odd number of bands, leading to a flat band in the 2D BZ, positioned at $E=0$. Since the 3D model can be seen as a stack of 2D Lieb lattices along any of the principal directions, the TPCs are protected by three coexisting chiral symmetries appearing on three mutually orthogonal planes. The chiral symmetries can be written as
\begin{equation}\label{eq:chiralsym}
 \Gamma_i H(\vec{k}_i) = - H(\vec{k}_i) \Gamma_i,
\end{equation}
with $i=x,y,z$. For the $k_x=\pi/2$ plane $\Gamma_x={\rm diag}(-1, -1, 1)$, on the $k_y=\pi/2$ plane we have $\Gamma_y={\rm diag}(1, -1, -1)$, whereas $\Gamma_z={\rm diag}(-1, 1, -1)$ at both $k_z=0$ and $k_z=\pi$. On each of these planes, the middle band of Eq.~\eqref{eq:Hlol} must be dispersionless and located at $E=0$. The two TPCs are found at points in which three orthogonal chiral planes meet (see Fig.~\ref{fig:system}, right panel) such that the product of three chiral symmetries $\Gamma_x\Gamma_y\Gamma_z=\mathds{1}$ must also anti-commute with the Hamiltonian. Hence, $H=0$ both for $\vec{k}=(\frac{\pi}{2}, \frac{\pi}{2}, 0)$ and $\vec{k}=(\frac{\pi}{2}, \frac{\pi}{2}, \pi)$ and the three chiral symmetries lead to triply degenerate bands at $E=0$ \cite{suppmat}. We note that, in previous models hosting TPCs, the latter are protected by lattice symmetries in the sense that, if TPCs exist, they are stable against any symmetry-preserving perturbation. In the model Eq.~\eqref{eq:Hlol} instead, the $\Gamma$ symmetries imply that the BZ \emph{necessarily} hosts TPCs. 
Finally, we observe that the double monopole charge of the TPCs leads to the existence of pairs of surface localized Fermi arcs $\hat{z}$ \cite{suppmat}.  

\textit{The physical setup}.
The potential leading to the lattice in Fig.~\ref{fig:system} and the artificial magnetic field can be obtained through a 3D generalization of the optical schemes applied for the simulation of the Hofstadter model \cite{Aidelsburger2013, Miyake2013}. The first element we consider is the optical lattice which defines the 3D Lieb lattice. It can be generated by three standing waves with the same frequency $\omega$, which are characterized by different linear polarizations $\hat{e}_i$ and must be red-detuned with respect to an atomic resonance to result in an attractive 3D lattice potential. The corresponding electric field reads:
\begin{equation}
\vec{E}(\vec{r},t) = E_0 \cos(\omega t) \sum_{j=x,y,z} \hat{e}_j \cos\left({\pi r_j}/{a}\right),
\end{equation}
where $a=\pi c/\omega$ is the lattice spacing, which we set to $1$ for convenience. In the case of orthogonal polarizations, the three lasers do not interfere with each other, producing a simple cubic lattice. We impose instead nonorthogonal polarizations given by $\hat{e}_x=(0,1,-1)/\sqrt{2}$, $\hat{e}_y=(-1,0,1)/\sqrt{2}$, $\hat{e}_z=(1,-1,0)/\sqrt{2}$, such that $\hat{e}_i \cdot \hat{e}_j =-1/2$ for $i\neq j$.
The optical lattice potential is affected by the interference of the three lasers and, after time averaging, reads:
\begin{multline} \label{potentialLOL}
V_{\rm ol}(x,y,z)=-\frac{E_0^2}{2}\left[\cos^2(\pi x) + \cos^2(\pi y) + \cos^2(\pi z) + \right. \\
\left. - \cos(\pi x)\cos(\pi y) - \cos(\pi x)\cos(\pi z) - \cos(\pi z)\cos(\pi y)\right]\,,
\end{multline}
where we have omitted an overall proportionality factor depending on the atomic polarizability.
This potential displays maxima in the (even, even, even) and (odd, odd, odd) cubic sublattices, where the attractive potential vanishes, whereas all the other sites have energy $-2E_0^2$, with barriers of height $E_0^2/2$ separating the nearest-neighbor sites \cite{suppmat}. 

The second element we engineer is the artificial gauge potential $A_z$. Following the standard approach for laser-assisted tunneling we must suppress the motion along the $\hat{z}$ direction and restore it through a pair of far-detuned Raman lasers \cite{Goldman2014}. As experimentally proved in Ref.~\cite{Miyake2013}, the gravitational potential $V_g(z) = mg z$ can be exploited for this purpose (possibly corrected by magnetic-field gradients) and the site-dependent tunneling in the $\hat{z}$ direction can then be reestablished through two Raman lasers at the same wavelength of the lattice, propagating along directions $\hat{x}$ and $\hat{z}$ with frequencies $\omega_1$ and $\omega_2$, respectively. These lasers determine a running-wave potential of the kind $W(\vec{r},t)=W_0 \cos(\omega_R t - \vec{k}_R\cdot \vec{r})$, where $\vec{k}_R \equiv \vec{k}_1-\vec{k}_2 \approx (\hat{x} + \hat{z}) \pi/a$ is the wave-vector difference between the Raman beams and $\omega_{R}=\omega_1 - \omega_2 \approx m g a/\hbar$ is the frequency difference, chosen to be resonant with the energy shift between adjacent sites.

In a regime such that the energy offset $mga$ is considerably larger than the bare tunneling amplitude $J$ [calculated without $V_g$ and $W(t)$] \cite{typvals}, the dynamics dictated by the time-dependent potential $V_{\rm ol} + V_g + W(t)$ can be approximated by an effective tight-binding Hamiltonian obtained in a rotating frame approximation \cite{Miyake2013,Aidelsburger2013},
\begin{multline} \label{ham2}
\hat H= \hspace{-1em} \sum_{\vec{r} \;{\rm s.t.}\; (-1)^{x+y}=-1} \hspace{-2em}-J_z\left[ (-1)^{x-z}c^\dag_{\vec{r} + \hat{z}}c_{\vec{r}} + {\rm H.c.}\right] +  \\
 -J\left[\sum_{\vec{r} \;{\rm s.t.}\; (-1)^{x+z}=-1} \hspace{-2em} c^\dag_{\vec{r} + \hat{y}}c_{\vec{r}} + 
 \sum_{\vec{r} \;{\rm s.t.}\; (-1)^{y+z}=-1}   \hspace{-2em} c^\dag_{\vec{r} + \hat{x}}c_{\vec{r}} + {\rm H.c.}\right]
\end{multline}
with $J_z = J \mathcal{J}_1\left(2W_0/(mga)\right)$ and $\mathcal{J}_1$ is a Bessel function. The above Hamiltonian is defined on a 3D Lieb lattice with the three-site unit cell and the Bravais vectors shown in Fig. \ref{fig:system} and corresponds to Eq.~\eqref{eq:Hlol} for $J_z = J$. In realistic systems $J_z<J$, and this introduces an anisotropy, 
which is not detrimental for the realization of the TPCs since it does not break any of the chiral symmetries.

\textit{Type-II crossings}.
Setting the Fermi level close to $E=0$, at positive or negative energies, leads to two disconnected electron or hole Fermi surfaces, one for each TPC. This is not the only possibility though as was initially understood for the simplest case of Weyl semimetals. One can distinguish between type-I Weyl points with a discrete Fermi surface and type-II Weyl cones, which are tilted by specific momentum-dependent perturbations such that the Fermi surface opens and the band-touching points lie at the boundary between electron and hole pockets \cite{Xu2015b, Soluyanov2015}. In the following we show that also TPCs may appear in different types due to a tilt of the zero-energy band.
By analogy with Weyl semimetals, we distinguish between the type-I TPCs of the Hamiltonian Eq.~\eqref{eq:Hlol} in which the middle band has a vanishing velocity along all momentum directions, and the novel type-II TPCs in which the middle band has a nonzero velocity at the band touching point.
This situation appears naturally when we generalize the Hamiltonian of the system by introducing an additional phase $\theta$,
\begin{widetext}
\begin{equation} \label{eq:Htype2}
 H_{\rm II} = - J\begin{pmatrix}
0 & 1 + e^{-i2k_x} & e^{i\left(k_z-k_x+k_y\right)+i\theta}-e^{-i\left(k_z+k_x-k_y\right)+i\theta}\\
1 + e^{i2k_x} & 0 & 1 + e^{i2k_y} \\
e^{-i\left(k_z-k_x+k_y\right)-i\theta}-e^{i\left(k_z+k_x-k_y\right)-i\theta} & 1 + e^{-i2k_y} & 0 
\end{pmatrix}.
\end{equation}
\end{widetext}
As a function of $\theta$ the middle band acquires a nonzero velocity as shown in Fig.~\ref{fig:tpcband}, right panel, but the chiral symmetries remain unbroken, such that the TPCs are still protected. Close to the band-touching points, the low-energy Hamiltonians to first order in $\theta$ at $k_x=k_y=\pi/2$ and $k_z=0,\pi$ are as follows:
\begin{equation}
 \frac{1}{2}H_{0,\pi} = \begin{pmatrix}
                     0 & ik_x & \pm ik_z \\
                 -ik_x &    0 &    -ik_y \\
              \mp ik_z & ik_y &        0 \\
             \end{pmatrix}\mp\theta
	     \begin{pmatrix}
              0 & 0 & k_z \\
              0 & 0 & 0 \\
              k_z & 0 & 0 \\
             \end{pmatrix}.
\end{equation}
In Weyl cones, the type-I crossing persists over a finite range of tilting angles before being converted into a type-II node. Here instead, even an infinitesimal value of $\theta$ leads to type-II TPCs. We find nonetheless that TPCs of type I remain protected as long as time-reversal symmetry is preserved as we specify in the Supplemental Material \cite{suppmat}.

The Hamiltonian $H_{\rm II}$ has a natural interpretation in terms of the artificial magnetic fluxes in the optical lattice. $\theta$ is a staggered phase in the vertical tunneling which depends on the the sublattice on the $xz$ plane. Therefore, in real space, such a system is obtained by correcting the $J_z$ term of the Hamiltonian \eqref{ham2} with
\begin{equation} \label{ham3}
\hat H_{\rm II,z}= -J_z \hspace{-2em} \sum_{\vec{r} \;{\rm s.t.}\; (-1)^{x+y}=-1} \hspace{-2em}\left[ e^{i\frac{\pi}{2}-i(-1)^{x-z}\left(\frac{\pi}{2}-\theta\right)}c^\dag_{\vec{r} + \hat{z}}c_{\vec{r}} + {\rm H.c.}\right]\,.
\end{equation}
If we consider the $xz$ plane and we embed the Lieb into a square lattice, such a magnetic flux configuration defines a checkerboard pattern with alternating $\pi \pm 2\theta$ fluxes obtained by the alternating $\pi -\theta$ and $\theta$ phases. 
To engineer these phases it is sufficient to add a new pair of counterpropagating Raman lasers oriented along the $\hat{z}$ axes such that, similar to the first pair of Raman lasers, their frequency difference is $\omega_R'=\omega'_1 - \omega'_2 \approx mga/\hbar$ and their wave-vector difference is $\vec{k}'_R = \hat{z}2\hbar \pi/a$ (we consider $\left|\omega_1 - \omega_1'\right|\gg \omega_R$ to avoid unwanted interferences with the first pair of Raman beams). By standard phase-lock techniques it is possible to shift the relative phase of these lasers by $\pi/2$ with respect to the first pair, thus defining the potential $W'(\vec{r},t)=W_1\cos(\omega'_R t - 2 \pi z/a-\pi/2)$.  The effect of this additional potential is to add a uniform imaginary contribution to the tunneling amplitude along $\hat{z}$ in \eqref{ham2}. Hence, the new amplitude reads $J_z e^{i\pi(x-z)} + i J_z' \equiv \mathcal{J}_z e^{i\phi (x-z)}$ with $\phi = \theta,\pi-\theta$ for $x-z$ even or odd, and $\theta = \arctan(J_z'/J_z)$ where $J_z'$ can be tuned through the Raman amplitude $W_1$ of the second pair of Raman lasers.

\emph{Conclusion}.
We have shown how to realize integer-spin analogs of Weyl cones in ultracold atomic systems. Due to the geometric constraints of the lattice, namely, the suppressed (even, even, even) and (odd, odd, odd) sites, the BZ shows robust symmetry-required TPCs. The latter can come in two types: the original spin-1 fermion described by a Hamiltonian $\vec{k}\cdot\vec{S}$, and a novel type-II crossing in which the middle band is tilted such that the TPC forms at the intersection between an electron and a hole pocket. Notice that this classification does not coincide with that of Ref.~\cite{Hu2017} in which different types of band crossings have a differing monopole charge. In our case, both the original and the tilted topological defects are double monopoles of the Berry curvature.

Our proposal only combines experimentally demonstrated building blocks, such as 3D optical lattices for ultracold fermions \cite{Schneider2008}, polarization-dependent potentials \cite{Grynberg1993,Mandel2003}, laser-induced gauge potentials \cite{Goldman2014,Aidelsburger2013,Miyake2013,Aidelsburger2014}, and 2D Lieb optical lattices \cite{Taie2015,Ozawa2017}. In these ultracold atom setups, the energy bands can be measured in a momentum-resolved manner using, for example, Bragg spectroscopy \cite{Clement2009,Ernst2009} or St\"uckelberg interferometry \cite{Kling2010}. Additionally, the triple band-touching points can be detected through Landau-Zener processes which measure the energy gaps by observing the non-adiabatic transitions between bands. Such techniques have been already applied for 2D systems \cite{Tarruell2012,Jotzu2014} and theoretically studied for 3D Weyl semimetals \cite{He2016}.
Furthermore, the band topology of the system could be probed by extending to three dimensions the techniques adopted to map the Berry curvature of the bands in planar geometries \cite{Duca2014,Flaschner2016}.

Finally, it is interesting to consider whether different types of topological objects can be obtained by tilting also in the case of higher-order band crossings, such as the spin-$\frac{3}{2}$ or spin-2 topological semimetals described in Ref.~\cite{Bradlyn2016}. The resulting anisotropy and the topologically protected crossing between electron and hole Fermi surfaces may lead to novel magneto-electric and transport properties \cite{Bovenzi2018, Pacholski2017}.

\emph{Acknowledgments}.
We warmly thank L. Duca, L. Lepori, A. Stern, and A. Trombettoni for fruitful discussions. L.F. and M.B. thank the Galileo Galilei Institute of Florence for hospitality during the development of this project. M.B. acknowledges support from the Villum Foundation. L.F. ackowledges funding from the European Research Council (ERC) under the EU's Horizon 2020 research and innovation programme (Grant Agreement No. 682629 - TOPSIM).

\bibliography{tpf}

\end{document}


\title{Supplemental Material: Geometrically protected triple-point crossings in an optical lattice}

\author{I. C. Fulga}
\affiliation{Institute for Theoretical Solid State Physics, IFW Dresden, 01171 Dresden, Germany.}

\author{L. Fallani}
\affiliation{Department of Physics and Astronomy, University of Florence \& LENS European Laboratory for Nonlinear Spectroscopy, 50019 Sesto Fiorentino, Italy}

\author{M. Burrello}
\affiliation{Niels Bohr International Academy and Center for Quantum Devices, University of Copenhagen, Juliane Maries Vej 30,
2100 Copenhagen, Denmark.}

\begin{abstract}
In this Supplemental Material we examine the symmetries of the 3D Lieb lattice, analyze the topologically protected Fermi arcs appearing on its surfaces, and we illustrate the optical lattice potential.
\end{abstract}

\maketitle

\section{Model symmetries}

To provide a more in-depth account of the symmetries of the 3D Lieb lattice, it is convenient to first write the Hamiltonian in terms of the momentum vectors introduced in the main text, $\vec{k}_1=(\frac{1}{2},0,-\frac{1}{2})$, $\vec{k}_2=(0, \frac{1}{2},-\frac{1}{2})$, and $\vec{k}_3=(0,0,1)$. In this basis, the Hamiltonian reads
\begin{equation}\label{eq:h123}
H = - J\begin{pmatrix}
0 & 1 + e^{-ik_1} & e^{-i\left(-k_3+k_1\right)+i\theta}-e^{-i\left(k_3-k_2\right)+i\theta}\\
1 + e^{ik_1} & 0 & 1 + e^{ik_2} \\
e^{i\left(-k_3+k_1\right)-i\theta}-e^{i\left(k_3-k_2\right)-i\theta} & 1 + e^{-ik_2} & 0 
\end{pmatrix}\,.
\end{equation}

\subsection{Time-reversal symmetry} 

When the phase $\theta$ is set to zero, all hopping terms are real, such that the Hamiltonian Eq.~\eqref{eq:h123} obeys time-reversal (TR) symmetry:
\begin{equation}
H(\vec{k}) = H^*(-\vec{k})\,.
\end{equation}

This symmetry distinguishes type-I from type-II TPCs, because the latter occur at time-reversal invariant momenta $(k_1,k_2,k_3)=(\pi,\pi,0)$ and $(\pi,\pi,\pi)$, while time-reversal constrains states at $+\vec{k}$ and $-\vec{k}$ to have the same energy. As such, TR guarantees that the middle band will have a vanishing Fermi velocity in all momentum directions, preserving the type-I nature of the crossing. In order to obtain a TPC of the second type, it is therefore necessary to break TR, for instance by adding a non-zero $\theta$ as shown in the main text.

\subsection{Chiral symmetries} 

The positions of the four chiral planes protecting the TPCs of Eq.~\eqref{eq:h123} are given by $k_1=\pi$, $k_2=\pi$, and $2k_3=k_1+k_2$, respectively. In addition, since the 3D Lieb lattice can be seen as a cubic lattice in which all (even, even, even) and (odd, odd, odd) sites have been suppressed, the model also obeys a global chiral symmetry. In real space, the unitary operator $P = (-1)^{n_x+ n_y+ n_z}$ anti-commutes with the Hamiltonian, where $(n_x, n_y, n_z)$ is a triplet of integers indexing the position of each site along the $\hat{x}$, $\hat{y}$, and $\hat{z}$ directions. In momentum space, the constraint imposed by the global chiral symmetry reads
\begin{equation} \label{Psym}
H(k_1,k_2,k_3)= -\Gamma_z H(k_1,k_2,k_3+\pi) \Gamma_z\,,
\end{equation}
where $\Gamma_z = {\rm diag}(-1,1,-1)$ as defined in the main text. Notice how our choice of a three site unit cell means that the global chiral symmetry $P$ is implemented together with a shift in the momentum $k_3$, as is common for transformations which act differently on neighboring unit cells. Counter-intuitively, this implies that a non-zero phase $\theta$ preserves $P$, even if it tilts the central band. It is only by doubling the size of the unit cell such as to make the symmetry momentum independent that one recovers the expected $E\to -E$ relation implied by chiral symmetry. In our case, using the doubled unit cell leads to a BZ in which the two TPCs overlap, and adding a non-zero $\theta$ tilts the overlapping central bands in opposite directions, thus preserving the $E\to -E$ symmetry.

We stress however that the global chiral symmetry $P$ plays no role in the protection of either type-I or type-II TPCs. To show this, we will prove that $P$ is independent of the planar chiral symmetries $\Gamma_{x,y,z}$, meaning that:
\begin{enumerate}
\item \label{i} there exist perturbations which break $P$ without breaking $\Gamma_{x,y,z}$ and, consequently, without destroying the TPCs, and
\item \label{ii} there exist perturbations which preserve $P$ the but break at least one of the planar chiral symmetries, thus resulting in a splitting of a TPC into multiple Weyl points.
\end{enumerate}

To prove the first statement, it is sufficient to consider perturbations which vanish at the four chiral planes. For instance, the term
\begin{equation}
H_{{\rm pert}}(k_1,k_2,k_3) =  \sin(k_1)\sin(k_2)\sin(2k_3-k_1-k_2) \,{\rm diag}(1,1,1)
\end{equation}
vanishes in the required planes and does not fulfill the relation \eqref{Psym}, namely $H_{{\rm pert}}(k_1,k_2,k_3) \neq -H_{{\rm pert}}(k_1,k_2,k_3+\pi)$. We conclude that $H_{{\rm pert}}$ breaks $P$ without breaking the planar chiral symmetries. $H_{{\rm pert}}$ describes a particular combination of long-range tunneling terms in the Bravais lattice and does not destroy the TPCs, since it vanishes at both of their positions.

\begin{figure*}[tb]
 \centering
 \includegraphics[width=0.4\columnwidth]{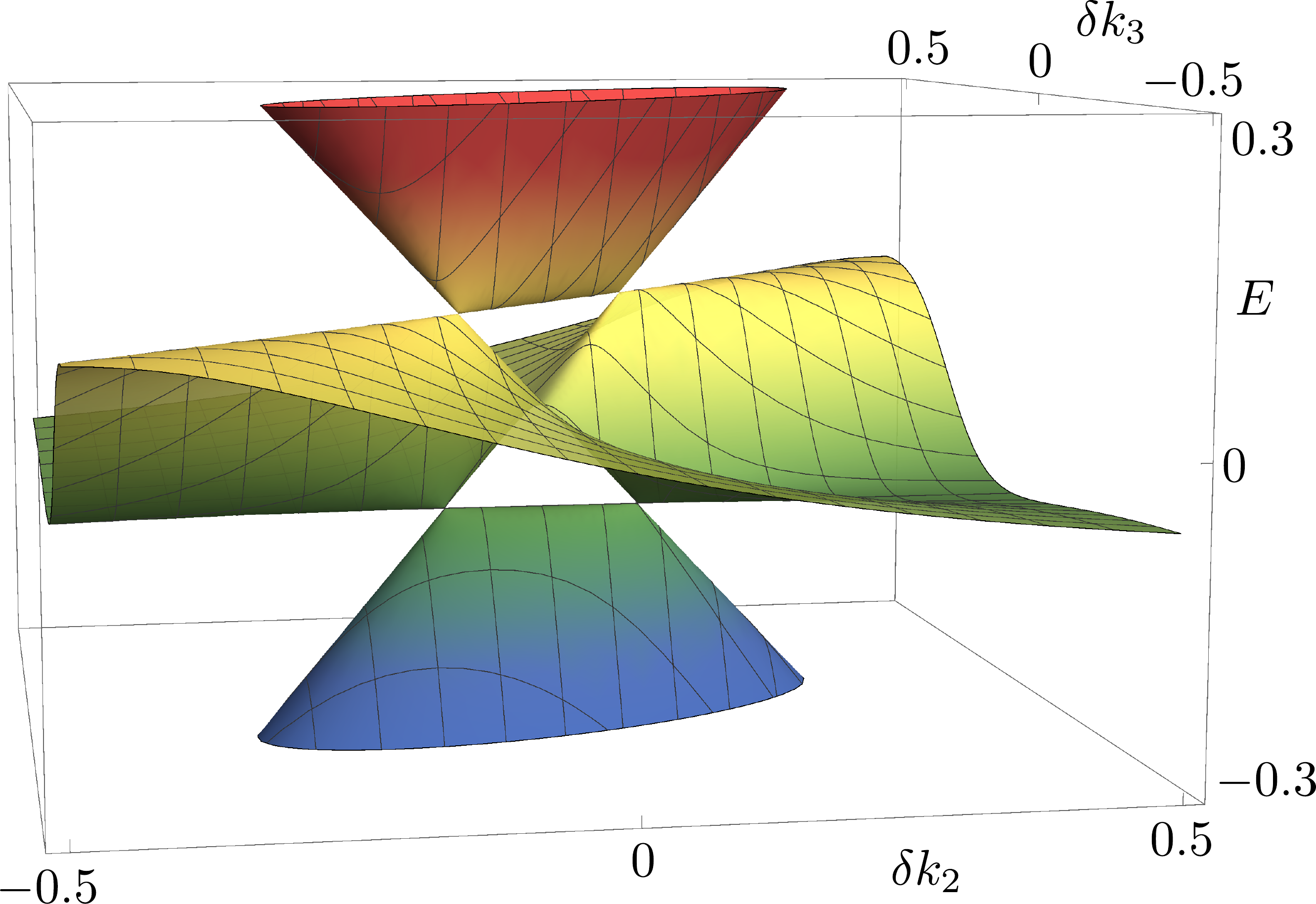}
 \caption{Bandstructure of the Hamiltonian Eq.~\eqref{eq:h123} with the addition of the perturbation Eq.~\eqref{eq:stagx}, using $J=1$ and $t_1=0.1$. The bands are shown at momenta $k_1=\pi$, $k_2=\pi+\delta k_2$, and $k_3=\pi+\delta k_3$. For nonzero $t_1$, the TPC splits into four Weyl cones of equal chirality. Each Weyl point has a shape corresponding to a phase transition between a type-I and a type-II Weyl node.\label{fig:splitting}}
 \end{figure*}

In order to prove the second statement, we show that the TPCs are not stable to perturbations which stagger the tunneling amplitudes along any direction of the lattice. This is a perturbation which does not affect $P$ but breaks one of the three planar chiral symmetries. Let us consider for example adding to Eq.~\eqref{eq:h123} the perturbation
\begin{equation}\label{eq:stagx}
H_{\rm{stag, \,x}} = - J\begin{pmatrix}
0 & t_1 & 0\\
t_1 & 0 & 0 \\
0 & 0 & 0 
\end{pmatrix}\,,
\end{equation}
which staggers the hoppings in the $\hat{x}$ direction and breaks the planar symmetry $\Gamma_x$. For $t_1\neq 0$, there is no point in momentum space in which the Hamiltonian completely vanishes and each of the TPCs splits into four Weyl points of the same chirality (see Fig.~\ref{fig:splitting}).
Two of them connect the bottom and middle bands, and the other two connect the middle band to the top one, as explained in Ref.~\onlinecite{Bradlyn2016}.

\section{Fermi arcs produced by the triple point crossings}

\begin{figure}[h]
 \includegraphics[width=0.6\columnwidth]{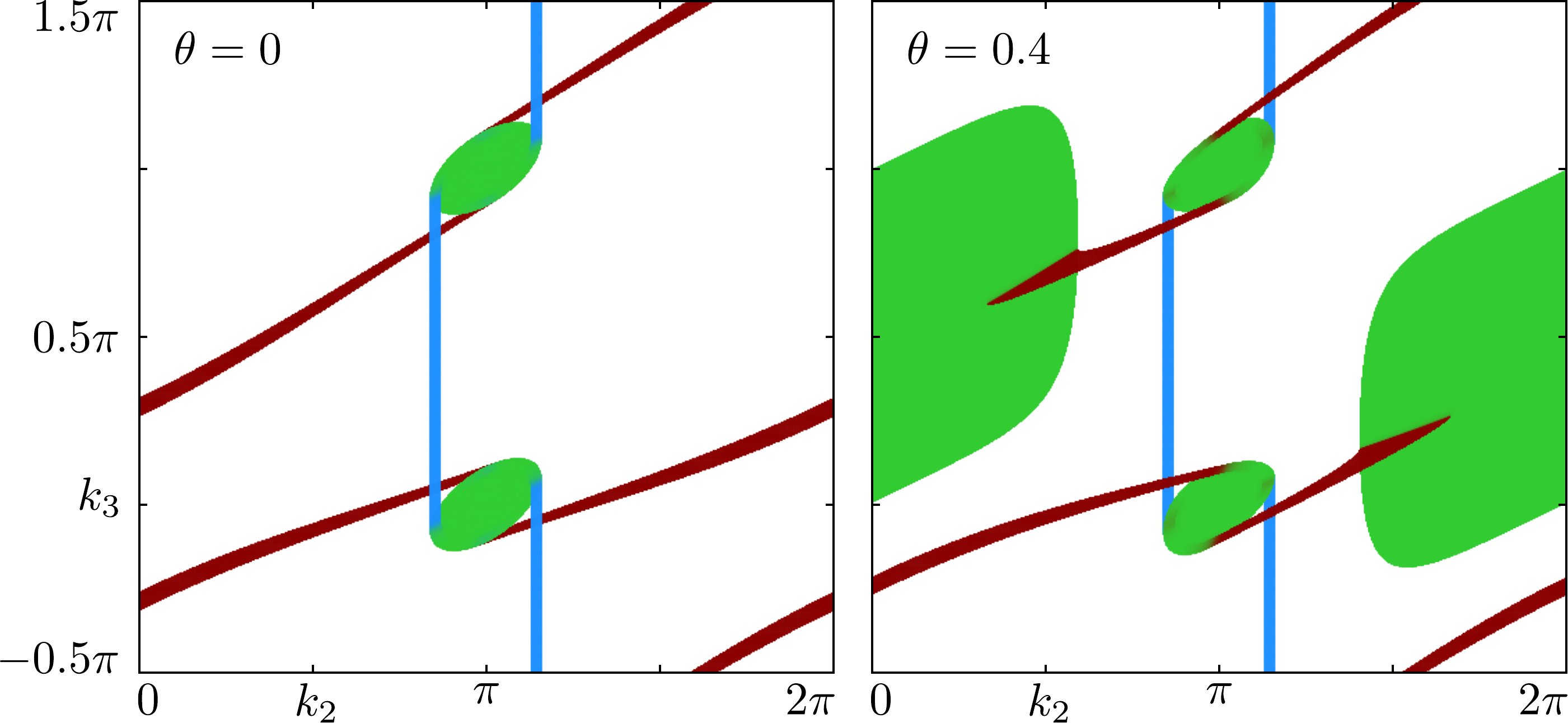}
 \caption{States with energy $0.4\leq E \leq 0.5$ are plotted in the surface BZ of an infinite slab of the model Eq.~(6). The system is infinite along $k_2$ and $k_3$, and consists of 100 unit cells in the $k_1$ direction. Bulk states are shown in green, while states on the two surfaces are plotted in blue and red, respectively. On each surface, two Fermi arcs originate from and connect the Fermi surfaces of the TPCs. In the left panel, $\theta=0$ corresponds to a type-I TPC, such that the central band is always at $E=0$. For the right panel, $\theta=0.4$ and the tilted central band of the type-II TPC leads to the appearance of a third Fermi surface, which hybridizes with the Fermi arcs.\label{fig:surfBZ}}
\end{figure}

The TPCs formed in the 3D Lieb lattice are chiral topological defects, characterized by a double monopole of the Berry curvature. As such, one expects to observe associated topologically protected surface states, the so-called Fermi arcs. To visualize them, we consider the TPC Hamiltonian Eq.~\eqref{eq:h123} in a slab geometry, infinite along $\vec{k}_2$ and $\vec{k}_3$, and consisting of 100 unit cells in the $\vec{k}_1$ direction (see Fig.~\ref{fig:surfBZ}). Two Fermi arcs appear on each of the surfaces of the slab, connecting the Fermi surfaces corresponding to the TPCs. To avoid the contribution of the bulk zero-energy states associated to the flat band, in Fig.~\ref{fig:surfBZ} we plot the states with energy in the range $0.4J\leq E \leq 0.5J$.

Interestingly, in this model the Fermi arc connectivity is such that surface states are present for any cut through the surface BZ. In the case of type-I TPCs (left panel), the central band remains flat, such that the only Fermi surfaces of the 2D BZ are those associated with the TPCs. Adding a tilting term $\theta=0.4$ (right panel) produces type-II TPCs with a tilted central band, resulting in the appearance of a third Fermi pocket in the surface BZ. The latter hybridizes with the Fermi arcs on one of the surfaces (red), but does not destroy their connectivity. Indeed, the Fermi arcs re-emerge from the other side of the third Fermi surface and show the same connectivity as for the type-I TPC, a signature of their preserved topological protection.

\section{Optical lattice potential}

\begin{figure*}[hb]
\centering
 \includegraphics[width=0.4\columnwidth]{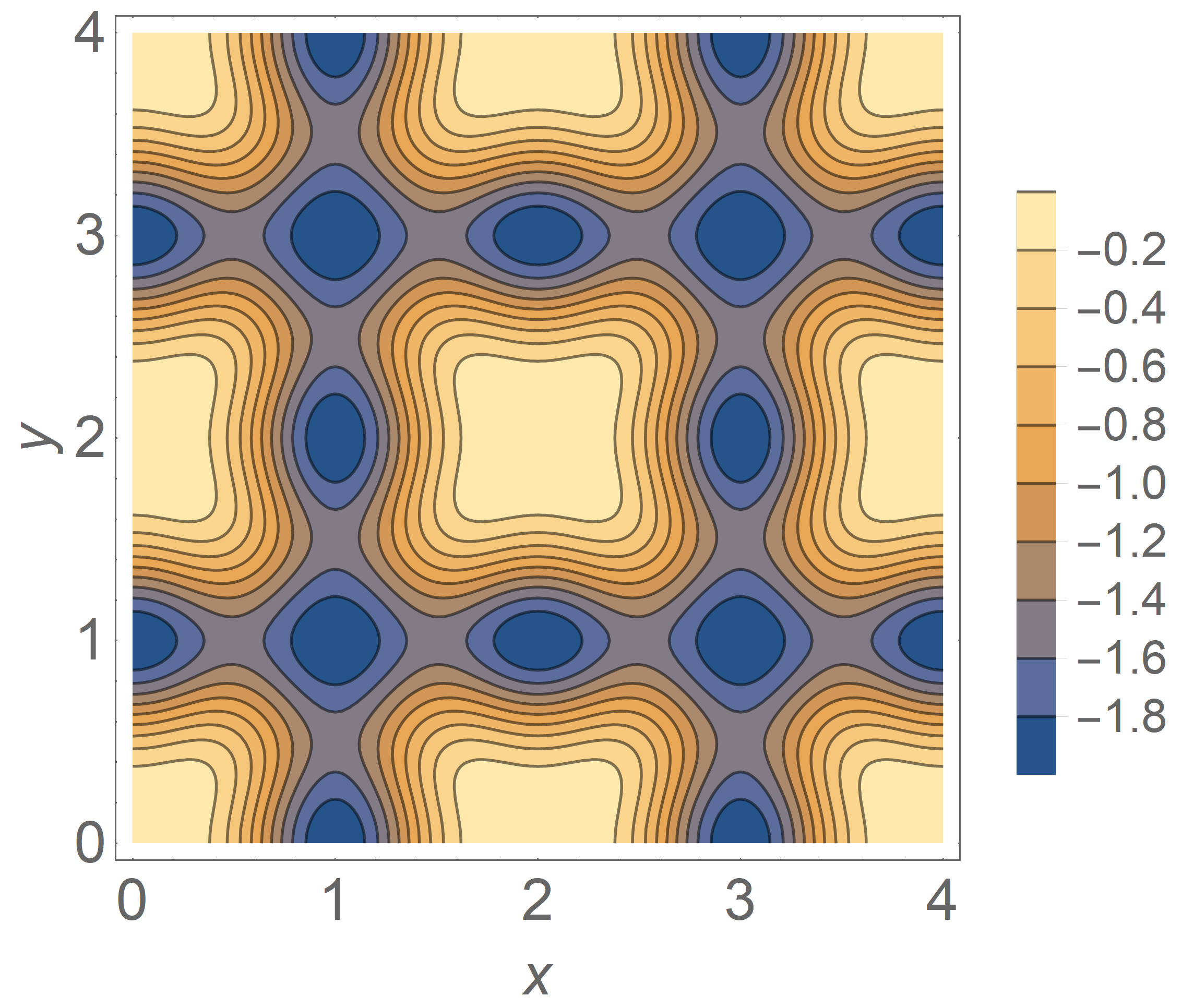} 
\hspace{0.03\columnwidth}
 \includegraphics[height=0.3\columnwidth]{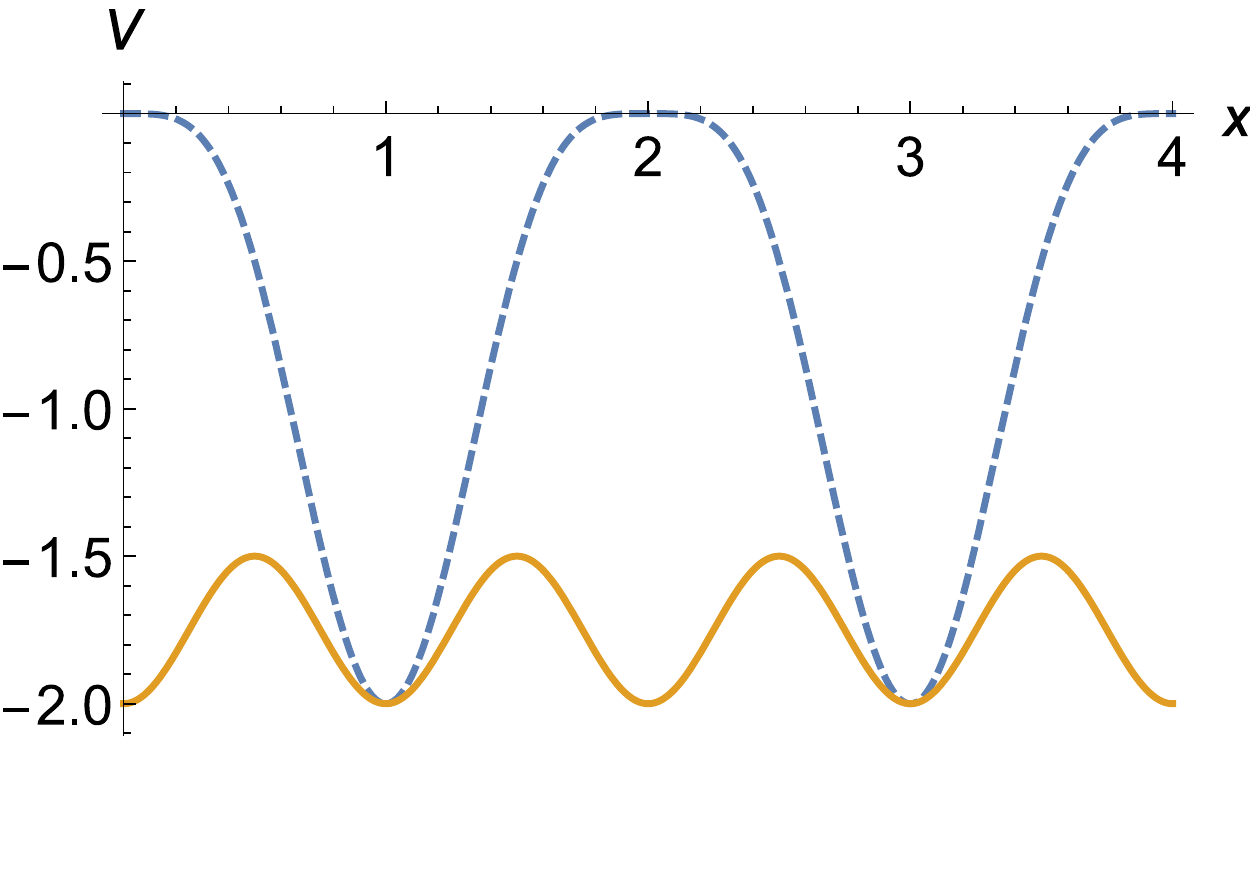}
\caption{Static optical lattice potential in Eq.~(4). Left: Contour plot of the potential in the plane $z=0$. The energies are in units of $E_0^2$ and the distances in units of the lattice spacing. The Lieb lattice potential can be clearly seen. The planes in the other directions are equivalent. Right: section of the potential Eq.~(4) for $y=z=0$ (blue dashed line) and $y=1,z=0$ (full orange line).}\label{fig:potential}
\end{figure*}

In Fig. \ref{fig:potential} we display the static optical lattice potential in Eq.~(4) of the main text.

\bibliography{tpf}